\documentclass[aps,prl,twocolumn,groupedaddress,showpacs]{revtex4}

\begin{document}


\title{Universal Measure of Entanglement}


\author{M. Hossein Partovi}
\affiliation{Department of Physics and Astronomy, California State University,
Sacramento, California 95819-6041}


\date{\today}

\begin{abstract}

A general framework is developed for separating classical and quantum
correlations in a multipartite system.  Entanglement is defined as the
difference in the correlation information encoded by the state of a system and
a suitably defined separable state with the same marginals. A generalization
of the Schmidt decomposition is developed to implement the separation of
correlations for any pure, multipartite state. The measure based on this
decomposition is a generalization of the entanglement of formation to
multipartite systems, provides an upper bound for the relative entropy of
entanglement, and is directly computable on pure states.  The example of pure
three-qubit states is analyzed in detail, and a classification based on
minimal, four-term decompositions is developed.

\end{abstract}


\maketitle


Entanglement is the simultaneous occurrence of superposition and correlation in composite systems, and gives
rise to some of the most counterintuitive phenomena of quantum mechanics. It was Einstein's early misgivings
about the nonlocality implicit in entangled states \cite{rosenfeld} that led to the celebrated EPR paper
\cite{EPR}.  This paper in turn inspired Schr\"{o}dinger's cat paradox and his prophetic identification of
entanglement as the most characteristic feature of quantum mechanics \cite{SCH}.  A striking confirmation of
this prediction is provided by Bell's renowned inequality, which is violated by a pure state if and only if it
is entangled \cite{Bell}. Long at the center of most foundational discussions, and widely regarded as the
essence of quantum strangeness, entanglement has in recent years emerged as the key {\it resource} in quantum
information processing \cite{Scr}. This remarkable transformation of paradox into paradigm was brought about by
the explosion of interest in quantum computing and communication in recent years \cite{CHN}.

The quantification of entanglement has thus emerged as a central problem in quantum information theory. Among
the proposed measures, the pioneering contributions of Bennett {\it et al.} \cite{Betal, Betal2} defined
entanglements of formation and distillation, ${E}_{F}$ and ${E}_{D}$ respectively, on considerations of
convertibility vis-\`{a}-vis maximally entangled pairs. Indeed ${E}_{F}$ and ${E}_{D}$ ($\leq {E}_{F}$) are in
effect benchmark {\it buy} and {\it sell} exchange rates for maximally entangled states, with any other ``good''
measure of entanglement expected to fall in between \cite{MHOR}.  On the other hand, the relative entropy of
entanglement \cite{vedraletal}, denoted by ${E}_{R}$, is based on distinguishability from the set of separable
states, with relative entropy serving as a measure of {\it contrast} \cite{contrast}.  These and most other
proposed measures are based on operational considerations of quantum communications or similarly motivated
mathematical axioms \cite{MHOR2}.  While bipartite entanglement is for the most part well understood, a general
formulation of the multipartite case is lacking and remains an outstanding unsolved problem.

In this Letter we develop a definition of entanglement based on
general, rather than operational or axiomatic, considerations.  We
also develop a generalization of the Schmidt decomposition
\cite{Schm} to implement this separation for pure, multipartite
states, and extend this to all multipartite states.  To see the
main idea in qualitative terms, let us consider a state $\rho$ of
a composite system together with its reduced (i.e., unipartite)
density matrices ${\rho}^{A}$, ${\rho}^{B}$, $\ldots$,
${\rho}^{Z}$, which represent the states of the components. The
latter are the quantum marginals of $\rho$, and collectively carry
less information than does $\rho$, the difference being the
correlation information encoded by $\rho$. Furthermore, because of
the superpositions it may contain, $\rho$ in general encodes more
correlation information than any classically correlated state that
possesses the same set of marginals. The difference, minimized
subject to suitable constraints, is then a measure of nonclassical
correlations, or entanglement, encoded by $\rho$.  Note that this
definition is based on separating the correlations encoded by a
quantum state into classical and nonclassical parts by reference
to a {\it common set of marginals} \cite{others}. Note also that
this notion of separation is already implicit in the standard
entanglement measure for pure bipartite states.

We now proceed to a quantitative formulation of the above ideas. Our objective is to define a classically
correlated state $\sigma$ associated with $\rho$. We require $\sigma$ to (i) be separable, i.e., a mixture of
pure, product states, and (ii) possess the same set of marginals as $\rho$. We further require $\sigma$ to
encode the greatest correlation information while having the least contrast \cite{contrast} against $\rho$.
Constraint (i) ensures that the correlations encoded by $\sigma$ are classical in nature \cite{W}, while
constraint (ii) guarantees that these correlations can be meaningfully compared to those of $\rho$. Furthermore,
the maximum correlation condition stipulates that $\sigma$ capture as much of the correlation content of $\rho$
as possible, so that only nonclassical correlations are counted as entanglement, while the minimum contrast
requirement requires that $\sigma$ differ as little as possible from $\rho$.  Following Ref. \cite{vedraletal},
we use relative entropy as our measure of contrast.

For pure bipartite states, minimizing relative entropy subject to
constraint (i) is sufficient to satisfy (ii) and the maximum
correlation condition as well \cite{vedraletal}. However, the
bipartite case is special, as evidenced by, e.g., the fact that
the Schmidt decomposition \cite{Schm} does not in general extend
beyond $N=2$ \cite{peres}. The distinguishing characteristic here
is the index ${N}_{ms}$, which we define as the number of distinct
marginal spectra possessed by a pure state \cite{Spec}. Now
${N}_{ms}=1$ for $N=2$, and $1 \leq {N}_{ms} \leq N$ for $N \geq
3$. Furthermore, one can easily show that {\it a pure state is
Schmidt-decomposable if and only if} ${N}_{ms}=1$ \cite{conv}.
Thus $N=2$ is special, and there is no {\it a priori} reason to
expect the maximum correlation and minimum contrast requirements
to be equivalent in the multipartite case. Since this equivalency
is a prerequisite for the above optimization problem, we add the
condition ${\rm tr}[(\sigma-\rho )\log(\sigma)]=0$, labelled
(iii), which turns out to have an interesting interpretation as
seen below.

Let ${\cal M}(\rho)$ be the set of states that satisfy constraints
(i)-(iii).  Then the correlation information contained in $\rho$
and $\sigma \in {\cal M}(\rho)$ are respectively given by
${C}^{\rho}=S(\check{\rho})-S( \rho)$ and
${C}^{\sigma}=S(\check{\sigma})-S(\sigma)$, where $S(\chi)$ stands
for the von Neumann entropy of $\chi$, and $\check{\rho}
{=}\check{\sigma} \stackrel{\rm def}{=}{\rho}^{A}\otimes
{\rho}^{B}\otimes \ldots \otimes {\rho}^{Z}$ represents the joint,
uncorrelated state of the marginals of $\rho$ or $\sigma$. The
nonclassical correlation information is thus given by the
difference ${C}^{\rho}-{C}^{\sigma}=S(\sigma)-S(\rho)$. Since the
above extremal conditions require the simultaneous maximization of
${C}^{\sigma}$ and minimization of $S(\rho \| \sigma)$, we must
ensure that their sum, $S(\check{\sigma})-S(\sigma)+S(\rho \|
\sigma)$, is a fixed quantity. This is precisely what is achieved
by (iii), which reduces the sum to $S(\check{\rho})-S(\rho)$, thus
guaranteeing the desired equivalency. Moreover, for $\sigma \in
{\cal M}(\rho)$, (iii) is equivalent to
\begin{equation}
S(\rho \| \sigma)+S( \sigma \| \check{\rho})=S(\rho \| \check{\rho}),
\label{0.5}
\end{equation}
which stipulates an additivity condition on contrast in comparing $\rho$ to $\check{\rho}$ via $\sigma$.  Note
that inasmuch as contrast ($=$ relative entropy) plays the role of distance here, we may interpret (iii) as
requiring that $\sigma$ be on a straight contrast line from $\rho$ to $\check{\rho}$.

The entanglement measure is now defined as the minimum of nonclassical
correlations:
\begin{eqnarray}
{E}(\rho) &\stackrel{\rm def}{=}& {\min}_{\sigma \in \cal{M}(\rho)} \, \,
[S(\sigma)-S(\rho)] \nonumber \\
&=&{\min}_{\sigma \in \cal{M}(\rho)} \, \, S(\rho \| \, \sigma), \label{1}
\end{eqnarray}
and will simply be referred to as ``entanglement.''  Note that because of
(iii), $\cal{M}(\rho)$ is not convex, so that there may be multiple local
minima for (\ref{1}).  Note also that ${E}={E}_{F}={E}_{R}$ on pure bipartite
states, and because of (ii) and (iii), ${E} \geq {E}_{R}$ on all states. This
completes our discussion of the general framework and definition of
entanglement.

Next, we develop a generalization of the Schmidt decomposition \cite{Schmidt}
as a means of developing an entanglement measure on pure, multipartite states.
Specifically, our objective is a minimal, orthogonal decomposition of any pure
state $| \psi \rangle $ such that its decohered form $\sigma(\psi)$ obeys
conditions (i)-(iii); we shall refer to these as {\it compact} decompositions.
Compact decompositions are in effect continued Schmidt decompositions: start
by applying the Schmidt decomposition to $| \psi \rangle$ with respect to $A$
and ($BC \ldots Z$) as the two subsystems,
\begin{equation}
| \psi \rangle ={ \sum}_{{i}_{a}}\sqrt{{\lambda}^{A}_{{i}_{a}}}| {\psi
}^{A}_{{i}_{a}}\rangle \otimes |{ \psi}^{BC \ldots Z} _{{i}_{a}}\rangle,
\label{1.5} \end{equation}
then to $|{ \psi}^{BC \ldots Z}_{{i}_{a}}\rangle$
with respect to $B$ and ($C \ldots Z$) as the two subsystems, and continue the
process to the last step where $Y$ and $Z$ are decomposed.  The result is
\begin{eqnarray}
|{\psi}
 \rangle={ \sum}_{{i}_{a}{i}_{b} \ldots {i}_{yz}}
{({\lambda}^{A}_{{i}_{a}}{\lambda}^{B}_{{i}_{a};{i}_{b}} \ldots
 {\lambda}^{YZ}_{{i}_{a}{i}_{b} \ldots {i}_{x};{i}_{yz}})}^{1 \over 2}
 {| {\psi}^{A} _{{i}_{a}}\rangle} \otimes \nonumber \\
{|{ \psi}^{B}_{{i}_{a};{i}_{b}} \rangle}\otimes \ldots \otimes {| {\psi}^{Y}
_{{i}_{a}{i}_{b}\ldots {i}_{x};{i}_{yz}} \rangle}\otimes {| {\psi}^{Z}
_{{i}_{a}{i}_{b}\ldots {i}_{x};{i}_{yz}}\rangle}. \label{2}
\end{eqnarray}
This expansion has a tree structure with each ket and its associated coefficient carrying a set of indices which
label the successive branch numbers. Accordingly, the collection of kets with all but the rightmost index (which
is set off by a semicolon) equal constitute an orthonormal set. For example,
\begin{equation}
\langle {\psi}^{W} _{{i}_{a}{i}_{b}\ldots {i}_{v};{i}_{w}} {| {\psi}^{W}
_{{i}_{a}{i}_{b}\ldots {i}_{v};{j}_{w}} \rangle}={\delta}_{{i}_{w}{j}_{w}}.
\label{2.5}
\end{equation}
For each such condition, there is a corresponding one for its Schmidt mate,
namely
\begin{equation}
\langle {\psi}^{XYZ} _{{i}_{a}{i}_{b}\ldots {i}_{v};{i}_{w}} {| {\psi}^{XYZ}
_{{i}_{a}{i}_{b}\ldots {i}_{v};{j}_{w}} \rangle}={\delta}_{{i}_{w}{j}_{w}}.
\label{2.6}
\end{equation}
Similarly, the collection of $\lambda$'s with all but the
rightmost index fixed constitutes a probability set. For example,
\begin{equation}
{\sum}_{{i}_{w}}{\lambda}^{W} _{{i}_{a}{i}_{b}\ldots {i}_{v};{i}_{w}}=1.
\label{2.7}
\end{equation}
Note that the last two subsystems in (\ref{2}) share a common set of indices
and a common coefficient as in a standard Schmidt decomposition. Thus there
are at most ${d}^{A}{d}^{B} \ldots {d}^{X}\min({d}^{Y},{d}^{Z})$ terms in the
above expansion, where, e.g., ${d}^{A}$ is the dimension of ${\cal H}^{A}$,
the Hilbert space belonging to subsystem $A$.   Except for the last pair,
changing the order of subsystems in Eq.~(\ref{2}) in general results in a
different decomposition. Therefore, allowing for possible degeneracies, there
are $N!/2$ compact decompositions for an $N$-partite system. For three qubits,
e.g., there are no more than three decompositions each containing no more than
4 terms.

The decohered form of $\psi$ in the basis of Eq.~(\ref{2}) is
\begin{eqnarray}
{\sigma}^{AB \ldots Z}(\psi)={ \sum}_{{i}_{a}{i}_{b} \ldots {i}_{yz}}
{{\lambda}^{A}_{{i}_{a}}{\lambda}^{B}_{{i}_{a};{i}_{b}} \ldots
 {\lambda}^{YZ}_{{i}_{a}{i}_{b} \ldots {i}_{x};{i}_{yz}}}
  {\pi}^{A}_{{i}_{a}} \otimes \nonumber \\
{ \pi}^{B} _{{i}_{a};{i}_{b}}\otimes \ldots \otimes  {\pi}^{Y}
_{{i}_{a}{i}_{b}\ldots {i}_{x};{i}_{yz}} \otimes
 {\pi}^{Z} _{{i}_{a}{i}_{b}\ldots {i}_{x};{i}_{yz}}, \label{3}
\end{eqnarray}
where each state vector in (\ref{2}) has been replaced with the
corresponding projection operator in (\ref{3}). The orthonormality
conditions stated above can now be used to show that (a) the
expansion given in Eq.~(\ref{3}) is orthogonal, (b) ${\sigma}^{AB
\ldots Z}(\psi)$ has the same marginals as $\psi$, and (c) ${\rm
tr} [{\sigma}^{AB \ldots Z}(\psi)- |\psi \rangle \langle \psi| ]
\log [{\sigma}^{AB \ldots Z}(\psi)] =0$. In other words,
${\sigma}^{AB \ldots Z}(\psi)$ satisfies conditions (i)-(iii)
above.  It is not difficult to show that among all possible ways
of decomposing a pure state by means of a sequence of Schmidt
decompositions, compact decompositions are unique in guaranteeing
conditions (a)-(c) stated above.

Denoting the set of $N!/2$ possible $\sigma $'s by ${\cal{M}}^{c}(\psi)$, we define the entanglement of $\psi$
following Eq.~(\ref{1}):
\begin{equation}
{E}^{c}(\psi)\stackrel{\rm def}{=} {\min}_{\sigma \in {\cal{M}}^{c}(\psi)} \,
\, S(\sigma), \label{4}
\end{equation}
where the superscript ``c'' refers to compact decompositions as the basis of
${E}^{c}$.  If this minimum is realized on, say, ${\sigma}^{AB \ldots
Z}(\psi)$, we have
\begin{eqnarray}
{E}^{c}(\psi)=-{\sum}_{{i}_{a}{i}_{b} \ldots {i}_{yz}}
{\lambda}^{A}_{{i}_{a}}{\lambda}^{B}_{{i}_{a};{i}_{b}} \ldots
{\lambda}^{YZ}_{{i}_{a}{i}_{b} \ldots {i}_{x};{i}_{yz}} \times \nonumber \\
\log({\lambda}^{A}_{{i}_{a}}{\lambda}^{B}_{{i}_{a};{i}_{b}} \ldots
{\lambda}^{YZ}_{{i}_{a}{i}_{b} \ldots {i}_{x};{i}_{yz}} ). \label{5}
\end{eqnarray}
The bounds on the entanglement of a system of $N$ parties each
with a Hilbert space of dimension $d$ are given by $0 \leq {E}^{c}
\leq (N-1) \log(d)$.  Note that Eqs.~(\ref{4}) and (\ref{5})
provide a readily computable, finite prescription for finding
${E}^{c}(\psi)$. This is in contrast to Eqs.~(\ref{1}) and
(\ref{9}) which involve nontrivial minimizations over infinite
sets.  It can also be verified that ${E}^{c}$ (as well as $E$) is
invariant under local unitary operations, and that its expected
value is non-increasing under local operations and classical
communications \cite{Betal2,vedraletal}.

Eq.~(\ref{5}) can be rearranged as
\begin{eqnarray}
{E}^{c}(\psi)=H({\lambda}^{A})+{\sum}_{{i}_{a}}
{\lambda}^{A}_{{i}_{a}}H({\lambda}^{B}_{{i}_{a};})+ \ldots +
\nonumber \\
{\sum}_{{i}_{a}{i}_{b} \ldots {i}_{x} }{\lambda}^{A}_{{i}_{a}}
{\lambda}^{B}_{{i}_{a};{i}_{b}}\ldots {\lambda}^{X}_{{i}_{a}{i}_{b} \ldots
;{i}_{x}} H({\lambda}^{YZ}_{{i}_{a}{i}_{b} \ldots {i}_{x};}), \label{5.1}
\end{eqnarray}
where, e.g.,
\begin{equation}
H({\lambda}^{B}_{{i}_{a};})\stackrel{\rm def}{=}-{\sum}_{{i}_{b}}
{\lambda}^{B}_{{i}_{a};{i}_{b}}\log({\lambda}^{B}_{{i}_{a};{i}_{b}})
\label{5.15}
\end{equation}
is the information associated with the probability set $\{ {\lambda}^{B}_{{i}_{a};} \}$.  This nested pattern is
of course inherited from the tree structure of compact decompositions noted above.

Next we consider the three-qubit case as an illustration of the foregoing construction. There are three compact
decompositions for a tripartite pure state $\psi$ which may be labelled ${\psi}^{ABC}$, ${\psi}^{BCA}$, and
${\psi}^{CAB}$, where, e.g.,
\begin{equation}
|{\psi}^{ABC} \rangle={ \sum}_{{i}_{a}{i}_{bc}}
{\lambda}^{A}_{{i}_{a}}{\lambda}^{BC}_{{i}_{a};{i}_{bc}} {| {\psi}^{A}
_{{i}_{a}}\rangle} \otimes {|{ \psi}^{B}_{{i}_{a};{i}_{bc}} \rangle} \otimes
{|{ \psi}^{C}_{{i}_{a};{i}_{bc}} \rangle}. \label{5.2}
\end{equation}
For three qubits represented as spin states, each of these can be brought to
the following standard form by local unitary transformations:
\begin{eqnarray}
{|{\psi}\rangle}^{ABC} = &\sqrt{{p}_{1}}& {|+ \rangle}^{A}\otimes {|+
\rangle}^{B}\otimes{|+\rangle}^{C} \nonumber \\
+& \sqrt{{p}_{2}}&{e}^{i{\alpha}} {|+ \rangle}^{A}\otimes {|-
\rangle}^{B}\otimes{|- \rangle}^{C} \nonumber \\
+& \sqrt{{p}_{3}}& {|- \rangle}^{A}\otimes {|{\theta}_{b}+
\rangle}^{B}\otimes{|{\theta}_{c}+ \rangle}^{C} \nonumber \\
+ &\sqrt{{p}_{4}}&{e}^{i{\beta}} {|- \rangle}^{A}\otimes {|{\theta}_{b} -
\rangle}^{B}\otimes{|{\theta}_{c} - \rangle}^{C}, \label{6}
\end{eqnarray}
where $|\theta + \rangle \stackrel{\rm def}{=}\cos (\theta/2)|+ \rangle + \sin
(\theta /2) |- \rangle$ and $|\theta - \rangle \stackrel{\rm def}{=}\sin
(\theta/2)|+ \rangle - \cos (\theta /2) |- \rangle$.  In addition to the
normalization condition ${\sum}_{i}{p}_{i}=1$, there are two further
constraints on the four real angles appearing in (\ref{6}),
\begin{equation}
\tan ({\theta}_{b}/2) \tan ({\theta}_{c}/2)=
 -{ {({p}_{1}{p}_{3})}^{1 \over 2}+ {({p}_{2}{p}_{4})}^{1 \over
2}{e}^{i({\beta}-{\alpha})} \over {({p}_{1}{p}_{4})}^{1 \over
2}{e}^{i{\beta}}+ {({p}_{2}{p}_{3})}^{1 \over 2}{e}^{-i{\alpha}} } \label{7},
\end{equation}
leaving a total of five independent, real parameters.  This is in agreement
with previous results \cite{Schmidt}, although these involve a minimum of five
terms in their decompositions as compared to four here.

When decohered, each of the three compact decompositions of $\psi$
yields a diagonal, separable density matrix of no more than four
terms which satisfies conditions (i)-(iii).  Among these, the
density matrix with the least entropy defines the entanglement of
$\psi$.  Therefore, ${E}^{c}(\psi)$ equals $
-{\sum}_{i}{p}^{*}_{i}\log({p}^{*}_{i})$, or simply $H({p}^{*})$,
where $\{ {p}^{*} \}$ is the probability set corresponding to the
minimal state. It is clear from this formula that $0 \leq
{E}^{c}(\psi) \leq \log(4)$. A simple example of a maximally
entangled three-qubit state is obtained from (\ref{6}) by setting
${\alpha}=0$, ${\beta}=\pi$, ${\theta}_{b}=0$, ${\theta}_{c}=\pi$,
and ${p}^{*}_{i}=1/4$. The three decompositions of this state are
identical due to its symmetry. This state may be represented as
\begin{eqnarray}
|{\psi}\rangle =& {1 \over 2} {|+ \rangle}^{A}\otimes {1 \over
2}{(|++ \rangle
+ |--\rangle )}^{BC} \nonumber \\
+& {1 \over 2} {|- \rangle}^{A}\otimes {1 \over 2}{(|+- \rangle +
|-+\rangle )}^{BC}. \label{8}
\end{eqnarray}
The entanglement of this state is $\log(4)$ units
which, by way of comparison, is twice as large as the entanglement
of the GHZ state ${1 \over 2} {( |+++ \rangle+ |--- \rangle)}$
\cite{GHZ}.  Referring to Eq.~(\ref{5.1}), we interpret this
result as representing $\log(2)$ units for the bipartite
entanglement of $A$ versus ($BC$), and $\log(2)$ units for the
average entanglement of the two Bell states that make up
${\rho}^{BC}$.

Pure three-qubit states are naturally classified in terms of the
ranks of their marginal spectra (see Ref. \cite{Schmidt} for an
alternative scheme). Class I corresponds to all marginals being of
rank one, and consists of product states with zero entanglement.
These correspond to ${p}_{i}=0$ for $i\neq1$ in (\ref{6}), with
${N}_{ms}=1$.  Class II corresponds to one marginal being of rank
one, the other two full rank. These are product states of pure
one-qubit and two-qubit states, with the entanglement carried by
the latter. Such states correspond to ${p}_{3}={p}_{4}=0$ in
(\ref{6}), with ${N}_{ms}=2$. Class III corresponds to all
marginals being full rank. This class is conveniently
subclassified as III-a,b,c, corresponding to ${N}_{ms}=1,2,3$,
respectively. With ${N}_{ms}=1$, Class III-a admits a standard
Schmidt decomposition \cite{conv} corresponding to
${p}_{2}={p}_{3}={\theta}_{b}={\theta}_{c}=0$ in (\ref{6}). The
further restriction ${p}_{1}={p}_{4}$ leads to the GHZ state
mentioned above. Class III-b has ${N}_{ms}=2$ and corresponds to
${\theta}_{b}={\theta}_{c}$ in (\ref{6}). Class III-c with
${N}_{ms}=3$ is the typical case shown in (\ref{6}). It is clear
that only class III states carry tripartite entanglement, while I
and II are better described as products of one-qubit and two-qubit
states.

Having described compact decompositions and the associated
entanglement measure ${E}^{c}$ on pure, multipartite states, we
now proceed to extend the latter to mixed states.  The method we
use was pioneered in Ref. \cite{Betal2}, and has since been found
to be a generally valid strategy for extending pure state measures
to all states \cite{UHLMANN}.  The key idea is to define the
entanglement of an ensemble of pure states to be the ensemble
average of the entanglement.  The minimum of this quantity over
all possible ensemble representations of a mixed state is then
defined as its entanglement. In symbols,
\begin{equation}
{E}^{c}(\rho)\stackrel{\rm def}{=}{\min}{\sum}_{\alpha} {p}_{\alpha}
{E}^{c}(|{\psi}_{\alpha} \rangle \langle {\psi}_{\alpha} |),  \label{9}
\end{equation}
where the minimum is taken over all possible ensemble
representations ${\sum}_{\alpha} {p}_{\alpha} |{\psi}_{\alpha}
\rangle \langle {\psi}_{\alpha} |$ of $\rho$. It is interesting to
note that ensemble averages of pure state entanglement are
implicit in the tree structure of compact decompositions, as may
be seen in Eq.~(\ref{5.1}), and in the above example of a
maximally entangled three-qubit state as well. Note also that
since ${E}^{c}$ and ${E}_{F}$ are equal on pure bipartite states,
they must coincide on all bipartite states. Therefore, ${E}^{c}$
as extended by Eq.~(\ref{9}) is a generalization of ${E}_{F}$ to
all multipartite states. Furthermore, because compact
decompositions satisfy conditions (ii) and (iii), we have ${E}^{c}
\geq {E}_{R}$ on all multipartite states.

Thus far we have defined two measures of entanglement, $E$ in
Eq.~(\ref{1}), and ${E}^{c}$ in Eqs.~(\ref{4}) and (\ref{9}).
Considering pure states for the moment, we defined these measures
by minimizing nonclassical correlations over ${\cal{M}}$ and
${\cal{M}}^{c}$, sets of states that satisfy conditions (i)-(iii).
While ${\cal{M}}$ includes all such states and is infinite,
${\cal{M}}^{c}$ includes states derived from compact
decompositions only and is finite. Thus $E \leq {E}^{c}$ on pure
states, with equality known to hold on bipartite states. Since we
expect compact decompositions to be minimal representations, it is
likely that ${E}^{c}={E}$ on all pure states, although we have not
succeeded in establishing this equality in general.  In the event
that the equality holds on pure states, there arises the question
of whether the two measures coincide on mixed states as well, also
an open question at present. If $E$ turns out to be strictly
smaller than ${E}^{c}$ on pure states, then ${E}^{c}$ provides a
readily computable upper bound for $E$.

Entanglement has been defined here as nonclassical correlation information,
without any direct reference to operational considerations \cite{poorly}. The
results, however, are remarkably close to operationally inspired measures of
entanglement, affirming the expectation that there must be a universal meaning
to the quantity of entanglement embodied in a quantum system.

This work was supported in part by a research grant from California State
University, Sacramento.

{}


\begin{thebibliography}{}
\bibitem{contrast} We use ``contrast,'' in contradistinction to ``distance,''
to designate any suitable measure of distinguishability.
\bibitem{Spec} To avoid trivial degeneracies, we will consider all spectra arranged
in a nonincreasing order.
\bibitem{conv} The converse states that, given a probability set, there is a
Schmidt-decomposable $N$-partite state in a suitably large Hilbert space all
of whose marginal spectra equal the given set.
\bibitem{poorly} The convertibility of multipartite to bipartite entanglement is
not well understood at present.
\end{thebibliography}
\end{document}